%% file: l4dc2025-sample.tex
\documentclass{l4dc2025}

\title[TamedPUMA: safe and stable IL with geometric fabrics]{TamedPUMA: 
safe and stable imitation learning\\with geometric fabrics}
\usepackage{times}
\usepackage{todonotes}
\newcommand{\SB}[1]{\textcolor{black}{#1}} 

\author{%
 \Name{Saray Bakker} \Email{s.bakker-7@tudelft.nl}\\
 \addr Department of Mechanical Engineering, TU Delft, The Netherlands
 \AND
 \Name{Rodrigo P\'{e}rez-Dattari} \Email{r.j.perezdattari@tudelft.nl}\\
 \addr Department of Mechanical Engineering, TU Delft, The Netherlands
 \AND
 \Name{Cosimo {Della Santina}} \Email{c.dellasantina@tudelft.nl}\\
 \addr Department of Mechanical Engineering, TU Delft, The Netherlands
 \AND
 \Name{Wendelin B\"{o}hmer}\Email{j.w.bohmer@tudelft.nl}\\
 \addr Department of Electrical Engineering, Mathematics \& Computer Science, TU Delft, The Netherlands
 \AND
 \Name{Javier Alonso-Mora}\Email{j.alonsomora@tudelft.nl}\\
 \addr Department of Mechanical Engineering, TU Delft, The Netherlands
}
\input{commands.tex}
\begin{document}

\maketitle

\input{sections/0_abstract}
\input{sections/1_introduction}
\input{sections/2_related_works}
\input{sections/3_preliminaries}
\input{sections/4_methods}

\input{sections/5_results}
\input{sections/6_conclusion}

\newpage
\acks{This project has received funding from the European Union through ERC, INTERACT, under Grant 101041863, and by the NXTGEN national program. Views and opinions expressed are however those of the authors only and do not necessarily reflect those of the European
Union or the NXTGEN national program. Neither the European Union nor the granting authority can be held responsible for them.}

\bibliography{references}

\end{document}

%% file: commands.tex
\renewcommand{\vec}{\boldsymbol}
\newcommand\X{\ensuremath{\mathcal{X}}}

\renewcommand\le{\ensuremath{\mathcal{L}_e}}

\newcommand\M{\ensuremath{\mat{M}}}

\newcommand\f{\ensuremath{\vec{f}}}

\newcommand\h{\ensuremath{\vec{h}}}

\newcommand\Spec{\ensuremath{\mathcal{S}}}

\newcommand\spec{\ensuremath{\left(\M,\vec{\xi}\right)_{\X}}}
\newcommand\x{\ensuremath{\vec{x}}}
\newcommand\xdot{\ensuremath{\dot{\x}}}
\newcommand\xddot{\ensuremath{\ddot{\x}}}

\newcommand\q{\ensuremath{\vec{q}}}

\newcommand\qdot{\ensuremath{\dot{\q}}}
\newcommand\qddot{\ensuremath{\ddot{\q}}}

\newcommand\J{\ensuremath{\mat{J}_{\phi_j}}}
\newcommand\Jt{\ensuremath{\mat{J}^T_{\phi_j}}}
\newcommand\Jdot{\ensuremath{\dot{\mat{J}}_{\phi_j}}}

\newcommand{\map}{\ensuremath{\phi}}

\newcommand{\norm}[1]{\left\lVert#1\right\rVert}
\newcommand\mat[1]{\ensuremath{\bm{#1}}}
\renewcommand\vec[1]{\ensuremath{\bm{#1}}}

\newcommand{\fTL}{\f^{\mathcal{T} \rightarrow \mathcal{L}}_{\theta}}
\newcommand{\fT}{\f^{\mathcal{T}}_{\theta}}

\newcommand{\fC}{\f^{\mathcal{C}}_{\theta}}

\acrodef{smp}[SMP]{Stable Motion Primitives}
\acrodef{gf}[GF]{Geometric Fabrics}
\acrodef{fpm}[FPM]{Forcing Policy Method}
\acrodef{cpm}[CPM]{Compatible Potential Method}
\acrodef{dnn}[DNN]{deep neural network}
\acrodef{il}[IL]{Imitation Learning}
\acrodef{ik}[IK]{Inverse Kinematics}
\acrodef{dmp}[DMP]{Dynamical Movement Primitives}
\acrodef{mpc}[MPC]{Model Predictive Control}
\acrodef{rl}[RL]{Reinforcement learning}
\acrodef{puma}[PUMA]{Policy via neUral Metric leArning}
\acrodef{fk}[FK]{forward kinematics}

%% file: sections/0_abstract.tex
\begin{abstract}
Using the language of dynamical systems, Imitation learning (IL) provides an intuitive and effective way of teaching stable task-space motions to robots with goal convergence. Yet, IL techniques are affected by serious limitations when it comes to ensuring safety and fulfillment of physical constraints. With this work, we solve this challenge via TamedPUMA, an IL algorithm augmented with a recent development in motion \SB{generation} called geometric fabrics. 
As both the IL policy and geometric fabrics describe motions as artificial second-order dynamical systems, we propose two variations where IL provides a navigation policy for geometric fabrics. 
The result is a stable imitation learning strategy within which we can seamlessly blend geometrical constraints like collision avoidance and joint limits.
Beyond providing a theoretical analysis, we demonstrate TamedPUMA with simulated and real-world tasks, including a 7-DoF manipulator. \href{https://autonomousrobots.nl/paper_websites/pumafabrics}{Link to website}
\end{abstract}

%% file: sections/1_introduction.tex
\section{Introduction}
As robotic solutions rapidly enter unstructured environments such as the agriculture sector and homes, there is a critical need for methods that allow non-experts to easily adapt robots for new tasks. Currently, experts manually program these tasks, a method that is costly and not scalable for widespread use. Moreover, these sectors demand that robots safely interact with dynamic environments where humans are present. 

A possible solution to this societal challenge comes from \ac{il}. Using this technique, robots can learn motion profiles from demonstrations provided by non-expert users. 
Furthermore, by encoding the learned trajectories as solutions of a dynamical system, established mathematical tools from dynamical system theory can be used to guarantee convergence to the task's goal state, as we discuss more in detail in Sec.~\ref{sec:rw}.
In a robotics context, these learned dynamical systems commonly encode the navigation policy towards a goal within a task space - such as the evolution of the end-effector pose of a manipulator while pouring water in a glass from all possible initial locations.
However, this focus on task space motions renders \ac{il} fundamentally limited when considering physical constraints involving the robot's body impacting itself or interacting with the external environment. 
Substantial recent research has looked into the problem but, as we discuss in Sec. \ref{sec:rw}, to the best of our knowledge, no \ac{il} method can simultaneously ensure stability and real-time fulfillment of physical constraints for systems with many degrees of freedom.

This work contributes to the state of the art by introducing \mbox{TamedPUMA}, a learning framework that builds on IL and geometric fabrics~\citep{ratliff2020optimization} to obtain inherently stable and safe motion primitives from demonstrations. 
By leveraging the recently introduced geometric framework for motion \SB{generation} called geometric fabrics, 
our approach learns stable motion profiles while considering online whole-body collision-avoidance and joint limits.
To make this possible, the learned task-space policy has to be formulated as a ${2^{\mathrm{nd}}\text{-order}}$ (neural) dynamical system as fabrics operate within the Finsler Geometry framework where vector fields must be defined at the acceleration level~\citep{bao2012introduction}. Also, the learned policy must admit an \textit{aligned} potential, which roughly means a scalar function whose gradient is aligned with the acceleration field when the velocity is null. In this paper, we formally assess how we can ensure that both conditions are met. The performance of our approach is evaluated with a simulated and real-world 7-DoF manipulator, where we also benchmark it against vanilla geometric fabrics, vanilla learned stable motion primitives and a modulation-based IL approach leveraging collision-aware IK.

%% file: sections/2_related_works.tex
\subsection{Related work}\label{sec:rw}
\textbf{Learning stable dynamical systems from demonstrations} Multiple methodologies have been introduced for learning stable dynamical systems~\citep{billard2022learning,hu2024fusion}. An influential methodology in learning stable dynamical systems is \ac{dmp} by~\cite{ijspeert2013dynamical}, ensuring convergence towards a simple manually-designed dynamical system. This approach is extended to non-Euclidean state spaces~\citep{ude2014orientation}, probabilistic environments~\citep{li2023prodmp, przystupa2023deepPMP}, and in the context of \ac{dnn}~\citep{pervez2017learning, ridge2020training}.
To ensure stability, several approaches enforce a specific structure on the function approximators, such as enforcing positive or negative definiteness~\citep{khansari2011learning,lemme2014neural,fernandez2018physically}, or invertibility~\citep{perrin2016fast, rana2020euclideanizing, urain2020imitationflow}.  
In contrast,~\cite{perez2023stable,perez2023deepmetric} enforce stability via additional loss functions derived using tools from the deep metric learning literature~\citep{kaya2019deep}. 
Importantly, ~\cite{perez2023deepmetric} extends the ideas of~\cite{perez2023stable} to more general scenarios, achieving better results in non-Euclidean state spaces and ${2^{\mathrm{nd}}\text{-order}}$ dynamical systems. These two features are essential for integrating such frameworks with fabrics, which, to the best of our knowledge, have not been simultaneously demonstrated in other works using time-invariant formulations, e.g.~\cite{rana2020euclideanizing, mohammadineural, saveriano2023learning}.

\noindent \textbf{From \ac{il} to whole-body motion \SB{generation}} For the safe operation of robots around humans, not only the dynamical system acting as the navigation policy should be stable but the whole-body motion is desired to be stable, collision-free, and within the system limits.
However, most existing \ac{il} solutions that incorporate obstacle avoidance when learning dynamical systems in end-effector space focus solely on avoiding collisions with the end-effector, e.g.~\cite{mohammadineural,ginesi2019DMPObstacle,huber2023avoidance}. 
One possibility to achieve full-body collision avoidance in such cases is by employing collision-aware \ac{ik} solvers~\citep{manipulationDrake, rakita2021collisionik}. Nevertheless, this ignores the desired acceleration or velocity profile specified by the learned dynamical system.

Another strategy is to combine \ac{il} with an optimization-based motion planner such as \ac{mpc}, inheriting the possibility to include kinodynamic and safety constraints and including the learned policy in the cost function. Several combinations exist, such as augmenting the cost function with a learned policy~\citep{xiao2022learning}, learning the complete cost function~\citep{tagliabue2023efficient}, and learning the complete MPC framework~\citep{ahn2023model}.
\cite{hu2023model} propose MPDMP$^+$ as a fast alternative for navigation amongst dynamic obstacles, combining \ac{dmp}s with \ac{mpc} where the cost function merges multiple demonstrations and includes obstacle avoidance using potential functions. 
Although some \ac{il}-\ac{mpc} frameworks can guarantee stability~\citep{ahn2023model, hewing2020learning}, where most methods lack these guarantees, their applications remain limited with no demonstrations in real-world robotics. 

\noindent \textbf{Geometric motion \SB{generation}} A fast alternative strategy to optimization-based motion planners lies in the field of geometric motion \SB{generation}. These techniques are inspired by Operational Space Control~\citep{khatib1987unified}, achieving stable and converging behavior for kinematically redundant robots using differential geometry~\citep{bullo2019geometric}. Recently, Riemannian Motion Policies and its extension, Geometric Fabrics, have been introduced~\citep{ratliff2018riemannian, ratliff2023fabrics}. In Geometric Fabrics, convergence is guaranteed with simple construction rules~\citep{ratliff2020optimization, Xie2020GF} and is also extended to dynamic environments~\citep{spahn2023dynamic}. These policies are shown to be beneficial in designing human-like motions~\citep{klein2022riemannian} and have a high planning frequency compared to optimization-based methods~\citep{spahn2023dynamic}.

\noindent \textbf{Combining geometric motion \SB{generation} and learning}
Geometric methods have been considered in conjunction with reinforcement learning~\citep{Wyk2024Geometric} and Bayesian learning~\citep{jaquier2020bayesian}. Most interestingly for the present work, \cite{xie2021imitation, xie2023neural} have looked into combining fabrics with \ac{il}. The method they propose is essentially different from TamedPUMA, as they directly learn the fabric, which results in limitations in terms of motion expressiveness.

%% file: sections/3_preliminaries.tex
\section{Preliminaries}
In this section, fundamental concepts are introduced for trajectory generation using artificial dynamical systems. Firstly, task and configuration spaces and their relations are specified in Section~\ref{sec: notations}, followed by an overview of motion \SB{generation} using geometric fabrics in Section~\ref{sec: geometric_fabrics}. For a detailed overview of geometric fabrics and differential geometries, we refer the reader to~\cite{ratliff2020optimization, ratliff2023fabrics}. Section~\ref{sec: stableMP} provides the necessary background on learned stable motion primitives and its stability properties.

\subsection{Configuration and task spaces} \label{sec: notations}
The configuration of the robot is denoted by $\q \in \mathcal{C}$ with its time derivatives $\qdot$ and $\qddot$. Here, $\mathcal{C}$ indicates the configuration space of the robot, which has a dimension of $n$. \emph{Tasks} can be defined in different task spaces $\mathcal{X}_j$. For instance, collision avoidance of a manipulator's end effector can be a task defined in the end-effector's space. Additionally, to address whole-body obstacle avoidance, multiple tasks can be defined for various coordinates along the robot's body. Furthermore, other tasks, such as reaching behaviors, can also be included. A task variable $\x_j \in \mathcal{X}_j$ denotes the value of the state representation for the $j$-th task space, where $j \in [M]$, $M$ denotes the number of task spaces, and $[M] = \{j \in \mathbb{Z}^{+}: j \leq M \}$. The relation between the configuration space and a given task space is stated via a twice-differential map $\map_j:\mathcal{C}\to\mathcal{X}_j$\SB{, and the map's Jacobian as $\vec{J}_{\phi_j}=\frac{\partial \map_j}{\partial \vec{q}}$}.

\subsection{Geometric fabrics} \label{sec: geometric_fabrics}
A dynamical system describes the behavior of a system using differential equations. In \emph{geometric fabrics}~\citep{ratliff2020optimization}, these dynamical systems describe an \emph{artificial} system generating desired trajectories for a robotic system. The desired motions are described using second-order nonlinear time-invariant dynamical systems, $\xddot = f(\x, \xdot)$.
In this framework, dynamical systems consist of two parts: one that conserves a \emph{Finsler energy} and another that is energy-decreasing. The energy-conservative part takes care of all \emph{avoidance tasks}, e.g., joint limit avoidance and obstacle avoidance, while the energy-decreasing part drives the system towards the goal. 

Let us first focus on the energy-conservative dynamical system. The behavior for an avoidance task is described using a second-order dynamical system, $\M(\x, \xdot)\xddot + \vec{\xi}(\x, \xdot) = \vec{0}$
within a task space $\X$ where $\M(\x, \xdot)$ is symmetric and invertible. If this system, often referred to as a spectral semi-spray $\vec{h}: \Spec = \spec$, is conserving a Finsler energy, it is called a fabric $\tilde{\vec{h}}: \Spec = (\tilde{\M}, \tilde{\vec{\xi}})_{\mathcal{X}}$~\citep{ratliff2020optimization}. The system $\vec{h}$ is transformed into a fabric $\tilde{\vec{h}}$ through \emph{energization} with the Finsler energy, i.e., ${\tilde{\vec{h}}=\mathrm{energize}_{\le}[\h(\x, \xdot)]}$~(Eq. 13 in \cite{ratliff2023fabrics}). Notably, the system's energization only changes the speed along the path, but not the path itself, if the system is homogeneous of Degree 2, i.e., $\h(\x, \alpha \xdot) = \alpha^2 \h(\x, \xdot), \ \forall \alpha \geq 0$.

Each avoidance task can be described by these energy-conservative fabrics, $\xddot_j = \tilde{\vec{h}}(\x_j, \xdot_j)$, within their own task space $\X_j$. To construct a whole-body policy for the robotic system, multiple task-space fabrics are mapped to the configuration space using a \emph{pullback operation} and summed. The pullback operation, 
\SB{$\mathrm{pull}_{\phi_{j}}\left(\tilde{\bm{M}}_j, \tilde{\vec{\xi}}_j \right)_{\mathcal{X}_j} = {\left(\Jt\tilde{\M}_j\J, \Jt(\tilde{\vec{\xi}}_j+\Jdot\qdot)\right)}_{\mathcal{C}}$,}
is constructed as a function of $\phi_{j}$, and maps the energy-conserving fabric to the configuration space, resulting in the system $\tilde{\h}_j(\q, \qdot): \Spec_j = (\tilde{\M}_j, \tilde{\vec{\xi}}_j)_{\mathcal{C}}$.
The \textit{specs} in \SB{$\mathcal{C}$} are summed, e.g. $(\Spec_1 + \Spec_2)_\mathcal{C} = (\tilde{\M}_1 + \tilde{\M}_2, \tilde{\vec{\xi}}_1 + \tilde{\vec{\xi}}_2)_\mathcal{C}$ where the resulting dynamical system $\qddot = \tilde{\vec{h}}(\q, \qdot)$ is a fabric as well, since the summation and pullback operations are closed under algebra. This combined fabric can be forced towards the minimum of a potential $\vec{\psi}(\q) \in \mathbb{R}$ and damped with a positive definite damping matrix\SB{~\citep{ratliff2020optimization}}. 
On a more abstract level, the desired dynamical system can be denoted as a combination of energy-conserving fabric $\tilde{\vec{h}}$ and an energy-decreasing navigation policy $\f$ which is damped,
\begin{equation} \label{eq: forced_fabric}
    \qddot = \tilde{\vec{h}}(\q, \qdot) + \f(\q, \qdot).
\end{equation}

\subsection{Learning stable motion primitives via PUMA} \label{sec: stableMP}
Similarly to geometric fabrics, \ac{puma} models a desired motion as a nonlinear time-invariant dynamical system. This method represents the dynamical system $\bm{f}^{\mathcal{T}}_{\theta}$ in one of the task spaces $\mathcal{X}_{j}$ (commonly the robot's end effector's space), denoted as $\mathcal{T}$, as a \ac{dnn} with weights $\theta$. The weights are optimized to imitate a set of demonstrations while ensuring convergence to a goal state $\x_{\text{g}} \in \mathcal{T}$~\citep{perez2023deepmetric, perez2023stable}. In other words, $\x_{\text{g}}$ must be a globally asymptotically stable equilibrium in the region of interest.

To enforce stability \SB{under minimization of the loss function}, a specialized loss is introduced and optimized alongside an imitation loss. To design this loss, it is necessary first to define a latent space $\mathcal{L}$ as the output of a hidden layer $l$ of the \ac{dnn}, such that 
\begin{equation} \label{eq: dnn_layers}
    \xddot = \bm{f}_{\theta}^{\mathcal{T}}(\x, \xdot) = \bm{\varphi}_{\theta}(\bm{\rho}_{\theta}(\x, \xdot)),
\end{equation}
where $\bm{\rho}_{\theta}:\mathcal{T} \to \mathcal{L}$ encodes the first $1, ..., l$ layers and $\bm{\varphi}_{\theta}:\mathcal{L} \to d\mathcal{T}$ the last $l+1, ..., L$ layers with $d\mathcal{T}$ representing the tangent bundle of $\mathcal{T}$, capturing the derivatives and tangent spaces associated with $\mathcal{T}$. Then, a \emph{mapped system} $\fTL(\x, \xdot)$ can be defined in $\mathcal{L}$ as 
\begin{equation}
    \ddot{\vec{y}} = \fTL(\x, \xdot) = \frac{\partial\bm{\rho}_{\theta}(\x, \xdot)}{\partial t}.
\end{equation}

These definitions enable the formulation of the \emph{stability conditions} introduced in \cite{perez2023stable}, which can be stated as:
\begin{theorem}[Stability conditions~\citep{perez2023deepmetric}]
\label{theo:puma_stability}
In the region $\mathcal{T}$, $\bm{x}_{\mathrm{g}}$ is a globally asymptotically stable equilibrium of $\bm{f}_{\theta}^{\mathcal{T}}$ if, $\forall \bm{x} \in \mathcal{T}$, (1) $\bm{y}_{\mathrm{g}}=\bm{\rho}_{\theta}(\x_{\mathrm{g}})$ is a globally asymptotically stable equilibrium of $\bm{f}_{\theta}^{\mathcal{T}\to\mathcal{L}}$, and (2) $\bm{\rho}_{\theta}(\x)=\bm{y}_{\mathrm{g}} \Rightarrow \bm{x}=\bm{x}_{\mathrm{g}}$.
\end{theorem}

These conditions can be enforced via the following loss~\citep{perez2023deepmetric}:
\begin{equation}
\label{eq:stable_loss}
    \ell_{\text{stable}} = \sum_{\bm{y}_{0}\in \mathcal{B}}\sum_{t \in \mathcal{H}} \max(0, m + d(\bm{y}_{\mathrm{g}}, \bm{y}(\bm{y}_{0}, t + \Delta t)) - d(\bm{y}_{\mathrm{g}}, \bm{y}(\bm{y}_{0}, t))),
\end{equation}
where $d(\cdot,\cdot)$ is a distance function, $m$ is a small margin hyperparameter, and $\bm{y}(\bm{y}_{0}, t)$ represents the value of $\bm{y}$ for an initial condition $\bm{y}_{0}$ and time $t$. Here, $\mathcal{B}$ is a batch of initial conditions obtained at every iteration by randomly sampling the workspace $\mathcal{T}$, and $\mathcal{H}$ contains different values of $t$ up to some time horizon $T$. To imitate the set of demonstrations, this method minimized the combined loss $\ell_{\textrm{PUMA}}=\ell_{\textrm{IL}} + \lambda \ell_{\textrm{stable}}$, where $\ell_{\textrm{IL}}$ is the behavioral cloning loss employed in \cite{perez2023stable} and $\lambda$ is a weight factor. 

%% file: sections/4_methods.tex
\section{TamedPUMA: Combining learned stable motion primitives and fabrics}
\label{sec:methods}
With learned stable motion primitives, complex tasks can be learned from demonstrations, while converging to the goal. In TamedPUMA, these learned dynamical systems are incorporated into the framework of geometric fabrics, generating stable and safe motions while respecting whole-body collision avoidance and physical constraints of the robot. In Section~\ref{sec: problem_formulation}, the problem formulation is discussed, followed by the two variations of TamedPUMA, the \ac{fpm} and \ac{cpm}, in Section~\ref{sec: gm} and ~\ref{sec: cm} respectively.

\subsection{Problem formulation} \label{sec: problem_formulation}
Consider there exists a dynamical system $\tilde{\f}$ that describes a desired motion profile for a robot, which induces the distribution of trajectories $\tilde{p}(\tau)$. For a given task variable $\x \in \mathcal{T}$, such as the robot's end-effector pose, we assume we can sample trajectories from this distribution. Then, if the dynamical system $\f^{\mathcal{T}}_{\theta}$ represents the evolution of the robot's state in $\mathcal{T}$, our objective is to minimize the distance between the distribution of trajectories induced by this system $p_{\theta}(\tau)$ and $\tilde{p}(\tau)$. Furthermore, $\f^{\mathcal{T}}_{\theta}$ must always reach a desired state $\x_{\mathrm{g}} \in \mathcal{T}$ while accounting for region avoidance constraints, namely, self-collisions, obstacles, and joint limits for M tasks defined in multiple task spaces $\mathcal{X}_{j}$. More formally, 
\begin{subequations}
\begin{align} \label{eq: problem_formulation_a}
    \ \f^{\mathcal{T}*}_{\theta} &= \underset{\f^{\mathcal{T}}_{\theta}}{\arg \min} \ D_{\text{KL}} \left( \tilde{p}(\tau) || p_{\theta}(\tau)) \right), \\
    \text{s.t.} & \ \ \
    \lim_{t\to\infty}||\x - \x_{\mathrm{g}}|| = 0, \ \ \  \forall \x \in \mathcal{T}, \label{eq: problem_formulation_b} \\ 
    & \ \ \ \x_j \in \mathcal{X}_{j}^{\text{free}}, \ \ \ \ \ \ \ \ \ \ \ \ \ \ \ \ \ \ \forall j \in [M]. \label{eq: problem_formulation_c}
\end{align}
\end{subequations}
This objective minimizes the Kullback-Leibler divergence between the estimated dynamical system and the target system~\eqref{eq: problem_formulation_a}. This divergence is often used as a distance measure between distributions as it is minimized through maximum likelihood estimation using distribution samples~\citep{osa2018algorithmic}. Moreover, the problem is subject to two constraints: (\ref{eq: problem_formulation_b}) the dynamical system should converge to a task-space goal $\x_{\mathrm{g}}$ in $\mathcal{T}$, and~\eqref{eq: problem_formulation_c} task-space states $\x_j, \forall j \in [M]$, should remain within the free space $\mathcal{X}_{j}^{\text{free}}$ where the dynamical system is well-defined, which could be the collision-free space or the space within the joint limits. 

From this formulation, \ac{puma} \emph{softly} addresses~\eqref{eq: problem_formulation_a} using the loss $\ell_{\textrm{IL}}$, while fabrics addresses~\eqref{eq: problem_formulation_c} through their geometric-aware formulation. Both approaches tackle the problem of stability; however, it remains challenging to combine both methods while ensuring convergence to the goal~\eqref{eq: problem_formulation_b}. In the following subsections, we propose two approaches to address this. 

\subsection{The Forcing Policy Method (FPM)} 
\label{sec: gm}
\begin{figure}[t]   
\hspace{\fill}
\includegraphics[width=\textwidth, keepaspectratio]{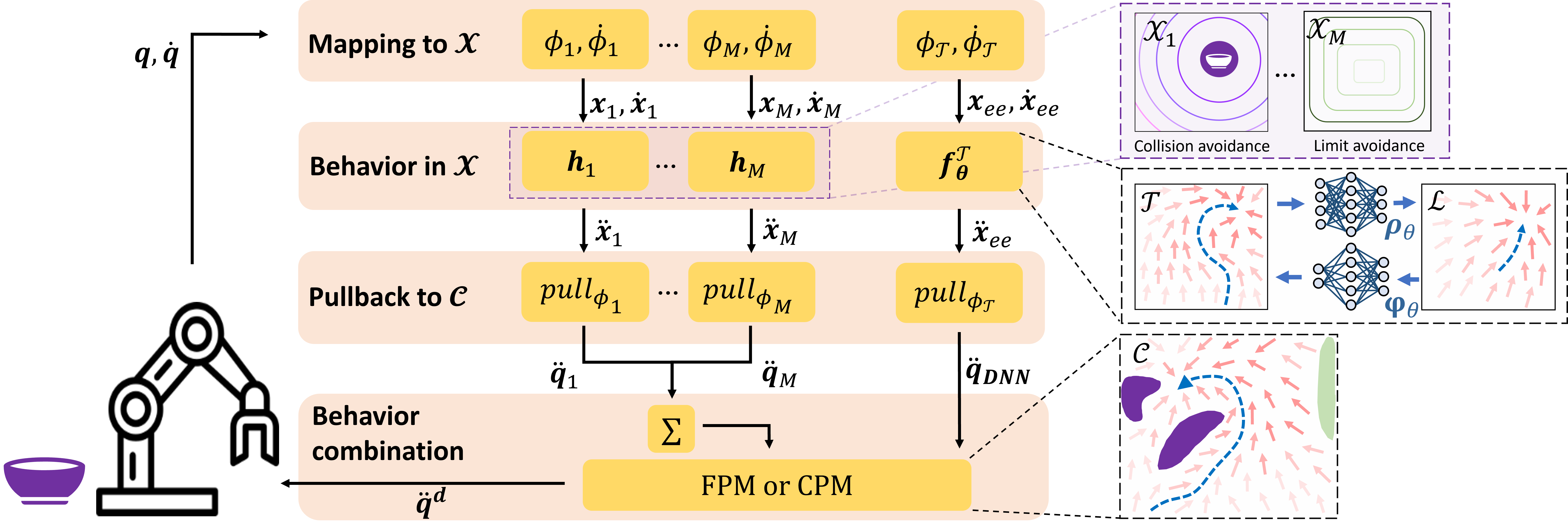}
\label{fig:illustration_potentials}
\hspace{\fill}
\vspace{-5mm}
\caption[]{\small{
This illustration of TamedPUMA shows the behavior design given the relationships between the different task and configuration-space variables. The joint angles and velocities get mapped into task space where the desired behavior is specified. Via fabrics, all avoidance behaviors are defined using the joint limits, and varying obstacle positions, e.g. the position of the bowl. The DNN captures the desired behavior of the end-effector position and orientation.
}}
\vspace{-4mm}
\label{fig: introduction_illustration}
\end{figure}

First, we introduce the \ac{fpm}. For this purpose, we define the dynamical system $\f^{\mathcal{C}}_{\theta}$ in configuration space resulting from applying a pullback operation, 
via \ac{puma}, in $\mathcal{T}$:
\begin{equation} \label{eq: fc_pulled}
    \qddot_{\text{DNN}} = \f^{\mathcal{C}}_{\theta}(\q, \qdot) = \mathrm{pull}_{\phi_{\mathcal{T}}}\left(\bm{f}^{\mathcal{T}}_{\theta}(\x_{\text{ee}}, \xdot_{\text{ee}})\right),
\end{equation}
where the task variable $\vec{x}_{\text{ee}}$ is the end-effector position and orientation of the robot as illustrated in Fig.~\ref{fig: introduction_illustration}.
Then, leveraging the definition of a forced system from Eq.~\eqref{eq: forced_fabric}, in the \ac{fpm} we propose to use the pulled system obtained via \ac{puma} as the forcing policy, 
\begin{equation} \label{eq: forced_by_NN}
    \qddot^{\text{d}} = \tilde{\vec{h}}(\q, \qdot) + \f^{\mathcal{C}}_{\theta}(\q, \qdot).
\end{equation}
Assuming $\ell_{\textrm{PUMA}}$ has already been minimized, the system $\fT$ comes to rest at $\x_{\mathrm{g}}$, implying that $\fC$ converges to $\q_{\mathrm{g}}$ where multiple values of $\q_{\mathrm{g}}$ may exist in the case of a redundant system. This collection of states $\q_{\mathrm{g}}$ corresponds to the \emph{zero set} of $\fC$. From Proposition II.17 in~\cite{ratliff2023fabrics}, we know that if the system in Eq.~\eqref{eq: forced_by_NN} reaches the zero set of $\fC$, it will stay there (which comes from the observation that fabrics are conservative). However, convergence of Eq.~\eqref{eq: forced_by_NN} to the zero set of $\fC$ is not formally assessed.
Re-evaluating the problem formulation suggests that the constraint in Eq~\eqref{eq: problem_formulation_b}, stating that the system should converge to the \SB{desired} minimum $\q_{\mathrm{g}}$, is therefore not guaranteed. In Sec.~\ref{sec: cm}, we propose a method with stronger convergence guarantees.

\subsection{The Compatible Potential Method (CPM)} \label{sec: cm}
As a second approach, we propose the \ac{cpm} \SB{leveraging} \emph{compatible potentials} to obtain a stronger notion of convergence. For a potential function compatible with a dynamical system, its negative gradient generally points in the same direction as the system's vector field. More formally:
\begin{definition}[Compatible potential~\citep{ratliff2023fabrics}]\label{def:compatible}
A potential function $\psi$ is \textit{compatible} with $\f$ if: (1) ${\partial \psi(\q) = \vec{0}}$ if and only if $\f(\q, \vec{0}) = \vec{0}$, and (2) $-\partial \psi^{\top} \f(\q, \vec{0}) > \vec{0}$ wherever $\f(\q, \vec{0}) \neq \vec{0}$.
\end{definition}

Building upon this, Theorem III.5 in \cite{ratliff2023fabrics} states that given a dynamical system $\f$ with a compatible potential, it is possible to construct another system that converges to the zero set of $\f$.
Crucially, this new system can incorporate the avoidance behaviors (e.g., joint limits and collision avoidance) in configuration space, represented by $\h(\q, \dot{\q}) = \sum_{j=1}^{M} \h_j(\q, \dot{\q})$, as:
\begin{equation} \label{eq: system_CPM}
    \qddot^{\text{d}} = \textrm{energize}_\mathcal{H}[\h+\f]+\gamma(\q,\qdot) \ \ \ \text{with} \ \ \ \gamma(\q, \qdot) = - \left(\frac{\qdot \qdot^{\top}}{\qdot^{\top}\vec{M}_{\mathcal{L}_e}\qdot}\right)\partial \psi - \beta \qdot.
\end{equation}
Here, we assume that the metric $\vec{M}_{\mathcal{L}_{e}}$ is bounded in a finite region, strictly positive definite everywhere and does not vanish or reduce rank if $\qdot \rightarrow \vec{0}$, $\beta>0$, and obstacle positions and joint limits are static. Consequently, we aim to leverage this result by using $\fC$ as the system $\f$ in Eq.~\eqref{eq: system_CPM} with the compatible potential. From the previous section, we already defined that the zero set of this system maps to the equilibrium $\x_{\mathrm{g}} \in \mathcal{T}$. Hence, it remains to find a compatible potential for this function to employ the result from Theorem III.5. 

Notably, if~\eqref{eq:stable_loss} is successfully minimized, it is possible to design a compatible potential for the system $\fT$ in the latent space $\mathcal{L}$ using the mapping $\bm{\rho}_{\theta}$ by setting $\xdot=\vec{0}$. Specifically, we can construct a potential using the latent-space variable $\vec{y}$ and the encoder $\vec{\rho}_{\theta}$ of PUMA as in Eq.~\eqref{eq: dnn_layers},
\begin{equation}
    \psi(\x) = \|\bm{y}_\mathrm{g} - \bm{y}\|^{2} =  \|\bm{\rho}_\theta(\x_{\mathrm{g}}, \vec{0}) - \bm{\rho}_\theta(\x, \vec{0})\|^{2}.
\end{equation}
To observe that this is a compatible potential of $\fT$, first, we highlight that since $\x_{\mathrm{g}}$ is asymptotically stable, we have ${\partial \psi(\x_{\mathrm{g}}) = \vec{0}}$ if and only if ${\f(\x_{\mathrm{g}}, \vec{0}) = \vec{0}}$. This satisfies the first condition of Definition~\ref{def:compatible}. Second, we note that for all $\x \neq \x_{\mathrm{g}}$, Equation~\eqref{eq:stable_loss} ensures that the value of $\psi(\x)$ decreases as $\fT$ evolves over time, except for some edge cases that could, in theory, occur if $\vec{\rho}_{\theta}$ has a large Lipschitz constant. These edge cases can be eliminated through regularization. Thus, this potential also satisfies the second condition of Definition~\ref{def:compatible} and therefore $\psi(\vec{x})$ is a compatible potential of $\vec{f}_{\theta}^\mathcal{T}$. Finally, for Eq.~\ref{eq: system_CPM} it only remains to express the gradient of this potential in configuration space, previously denoted as $\partial \psi$. For clarity, we will henceforth write this as $\partial \psi / \partial \q$. To achieve this, we require the forward kinematics from configuration space $\mathcal{C}$ to task space $\mathcal{T}$, denoted $\phi^\mathcal{T}$. Then, we obtain
\begin{equation} \label{eq: partial_potential}
\partial \psi = \frac{\partial \psi}{\partial \q} = \frac{\partial \psi}{\partial \x} \cdot J_{\phi^\mathcal{T}}(\x),
\end{equation}
where $J_{\phi^\mathcal{T}}$ is the Jacobian matrixof the forward kinematics, commonly available in robotic frameworks. The term $\partial \psi / \partial \x$ can be approximated via automatic differentiation tools for DNNs. 

\textbf{Assumptions on $\vec{M}_{\mathcal{L}_{e}}$} To conclude this section, it is relevant to discuss further the implications of the assumptions on $\vec{M}_{\mathcal{L}_{e}}$. 
For a fabric describing collision avoidance, two cases exist as the spec describing the fabric must be boundary conforming~\citep{ratliff2020optimization}. (1) The metric $\vec{M}_{\mathcal{L}_{e}}$ is finite along the eigen-directions parallel to the boundary's tangent space but goes to infinity along directions orthogonal to the tangent space. (2) The metric $\vec{M}_{\mathcal{L}_{e}}$ is a finite matrix along all trajectories, implying that $\vec{M}_{\mathcal{L}_{e}}$ is also finite in the limit when $t\rightarrow \infty$. Considering our previous assumptions on $\vec{M}_{\mathcal{L}e}$ for~\eqref{eq: system_CPM}, i.e. the metric is bounded in a finite region and strictly positive definite everywhere, we can design fabrics only according to the second scenario. This implies that, in the limit, we ensure convergence to the zero set of the forcing policy $\f^{\mathcal{C}}_{\theta}(\q, \qdot)$; however, collision avoidance is not guaranteed in the limit, as barrier-like functions that go to infinity on the boundary cannot be used to construct $\vec{M}_{\mathcal{L}_e}$.
For more details, we refer the reader to the attached material\footnotemark[1]{}.
\footnotetext[1]{\href{https://autonomousrobots.nl/paper_websites/pumafabrics}{Link to the attached material: https://autonomousrobots.nl/paper\_websites/pumafabrics}}

%% file: sections/5_results.tex
\section{Experimental Results} \label{sec: results}
In this section, we explore TamedPUMA's capabilities in constructing stable and collision-free motions for a 7-DOF manipulator. \SB{Details regarding the \ac{dnn}, differential mappings, a discussion on the limitations, code and videos are included in the attached material\footnotemark[1]{}.} 
\subsection{Experimental setup and performance metrics}
To show the performance of the two variations of TamedPUMA, \ac{fpm} and \ac{cpm}, simulations using the Pybullet physics simulation~\citep{coumans2019pybullet} and real-world experiments are performed on a 7-DoF KUKA iiwa manipulator. 
Two tasks are analyzed, picking a tomato from a crate and pouring liquid from a cup, where a \ac{dnn} is trained for each task using 10 demonstrations\SB{, recording end-effector positions, velocities and accelerations}. 
The proposed \ac{fpm} and \ac{cpm}, are compared against vanilla geometric fabrics, vanilla \ac{puma} and \emph{modulation-IK}.
Modulation-IK modifies PUMA to be obstacle-free within the task space using a modulation matrix~\citep{khansari2012dynamical}. Then, whole-body collision avoidance is achieved by tracking the modified desired pose using a collision-aware \ac{ik}~\citep{manipulationDrake}.
All methods are evaluated based on their \textit{success rate} and \textit{time-to-success}, which respectively indicate the ratio of successful scenarios and the time required for the robot to reach the goal pose. Successful scenarios ensure collision-free motions with an end pose satisfying $\norm{\x_{\text{ee}} - \x_{\text{g}}}_2 < 0.05 \ m$. 
In addition, the \textit{computation time} is denoted, and the \textit{path difference} to the desired path by PUMA as
$\frac{1}{P}\sum\norm{\vec{x}_{\text{ee}} - \vec{x}_{\text{PUMA}}}_2$
where $\vec{x}_{\text{ee}}$ and $\vec{x}_{\text{PUMA}}$ correspond to the end-effector poses along the path with length $P$ of the analyzed method and PUMA respectively. The path difference is computed in obstacle-free scenarios, where we aim to track the DNN as closely as possible, and in obstacle-rich allowing for deviations from this path.
\subsection{Simulation experiments on a 7-DOF manipulator}
\begin{table*}
    \footnotesize
	\begin{center}
 	\caption{\small{Statistics for 30 simulated scenarios. 
    The path difference to PUMA is measured in an obstacle-free environment, while all other metrics are compared in an obstacle-rich environment. 
    }}
    \label{tab: results_simulation}
		\begin{tabular}{c|c|c|c|c}
			  & Success-Rate & Time-to-Success [s] & Computation time [ms] & Path difference to PUMA \\
			\hline
			PUMA & 0.40 & 4.75 $\pm$ 1.31 &  4.41 $\pm$ 0.26 & 0  \\
			Modulation-IK & 0.57 & 5.75 $\pm$ 5.83 &  7.08 $\pm$ 5.98 & 0.39 $\pm$ 0.42 \\
			Fabrics & 0.83 & 8.18 $\pm$ 5.54 &  0.39 $\pm$ 0.03 & 0.22 $\pm$ 0.27 \\
			FPM (ours) & 1.00 & 6.97 $\pm$ 3.78 & 5.22 $\pm$ 0.22 & 0.02 $\pm$ 0.04\\
			CPM (ours) & 1.00 & 7.01 $\pm$ 3.57 & 6.55 $\pm$ 0.96 & 0.04 $\pm$ 0.06 \\
		\end{tabular}
	\end{center}
    \vspace{-4mm}
\end{table*}
In simulation, 30 realistic scenarios are explored, including 15 scenarios of a tomato-picking task and 15 scenarios of a pouring task. In each task, the initial robot configuration and obstacle locations change. Moreover, 10 scenarios included moving goals, and 7 scenarios included moving obstacles. 
For each scenario, at least one obstacle is situated between the initial configuration and the goal pose. 
The last five links on the robotic chain are considered for collision avoidance, given that the first three links have restricted movement due to the base being mounted to the table. The shape of the last five links is approximated using spheres with a radius of 9 $cm$, and 7 $cm$ for the final link. 

As depicted in~Table~\ref{tab: results_simulation}, the two TamedPUMA variations improve the success rate over PUMA by enabling whole-body obstacle avoidance. In contrast to geometric fabrics, \ac{fpm} and \ac{cpm} can track the desired motion profile leading to a smaller path difference with~\ac{puma} of 0.02 $\pm$ 0.04 and 0.04 $\pm$ 0.06 respectively, compared to geometric fabrics, 0.22 $\pm$ 0.27, in an obstacle-free environment.
In an obstacle-rich environment, geometric fabrics result in a deadlock in 6 of the 30 scenarios where the robot does not reach the goal as it is unable to move around the edge of the crate or object. 
The benchmark Modulation-IK is also unable to achieve all tasks due to collisions or deadlocks\SB{, as it cannot find a feasible solution converging towards the goal}. 
TamedPUMA thereby inherits the efficient scalability to multi-object environments from fabrics, as computation time in an environment with 1000 obstacles is only 5.9 $\pm$ 0.8 $ms$ and 7.0 $\pm$ 1.1 $ms$ for FPM and CPM respectively\SB{, a neglectable increase with respect to the two obstacles considered in Table~\ref{tab: results_simulation}}, while optimization-based methods like Modulation-IK scale poorly with large numbers of obstacles with a computation time of 3.5$\cdot10^3$ $\pm$ 10.7$\cdot10^3$ $ms$ in an environment with 1000 obstacles.
Although \ac{cpm} offers stronger theoretical guarantees than \ac{fpm}, performance is similar (see Table~\ref{tab: results_simulation}). Even though we do not optimize over the time-to-success, both \ac{fpm} and \ac{cpm} achieve the task within a reasonable time while remaining collision-free.
Computation times are 4-7 $ms$ on a standard laptop (i7-12700H) making the methodologies well suitable for real-time reactive \SB{motion generation} in dynamic environments. 
\vspace{-2mm}

\subsection{Real-world experiments on a 7-DOF manipulator} \label{sec: realworld}
\begin{figure*}[t]   
\subfigure[\footnotesize{Initial pose}]{\includegraphics[width=0.19\textwidth, keepaspectratio]{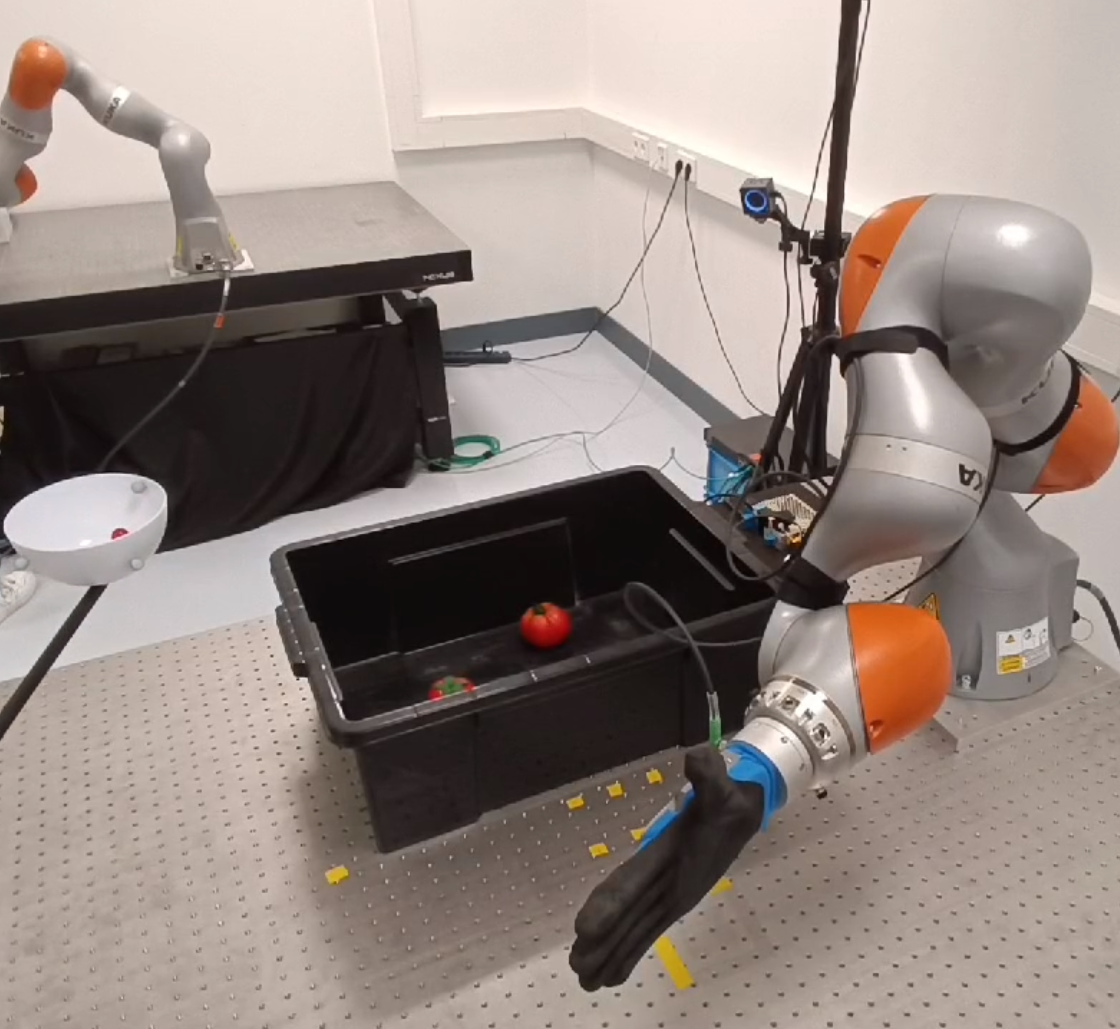}}\label{fig:subfig5}
\hspace{\fill}
\subfigure[\footnotesize{Bowl approaches}]{\includegraphics[width=0.19\textwidth, keepaspectratio]{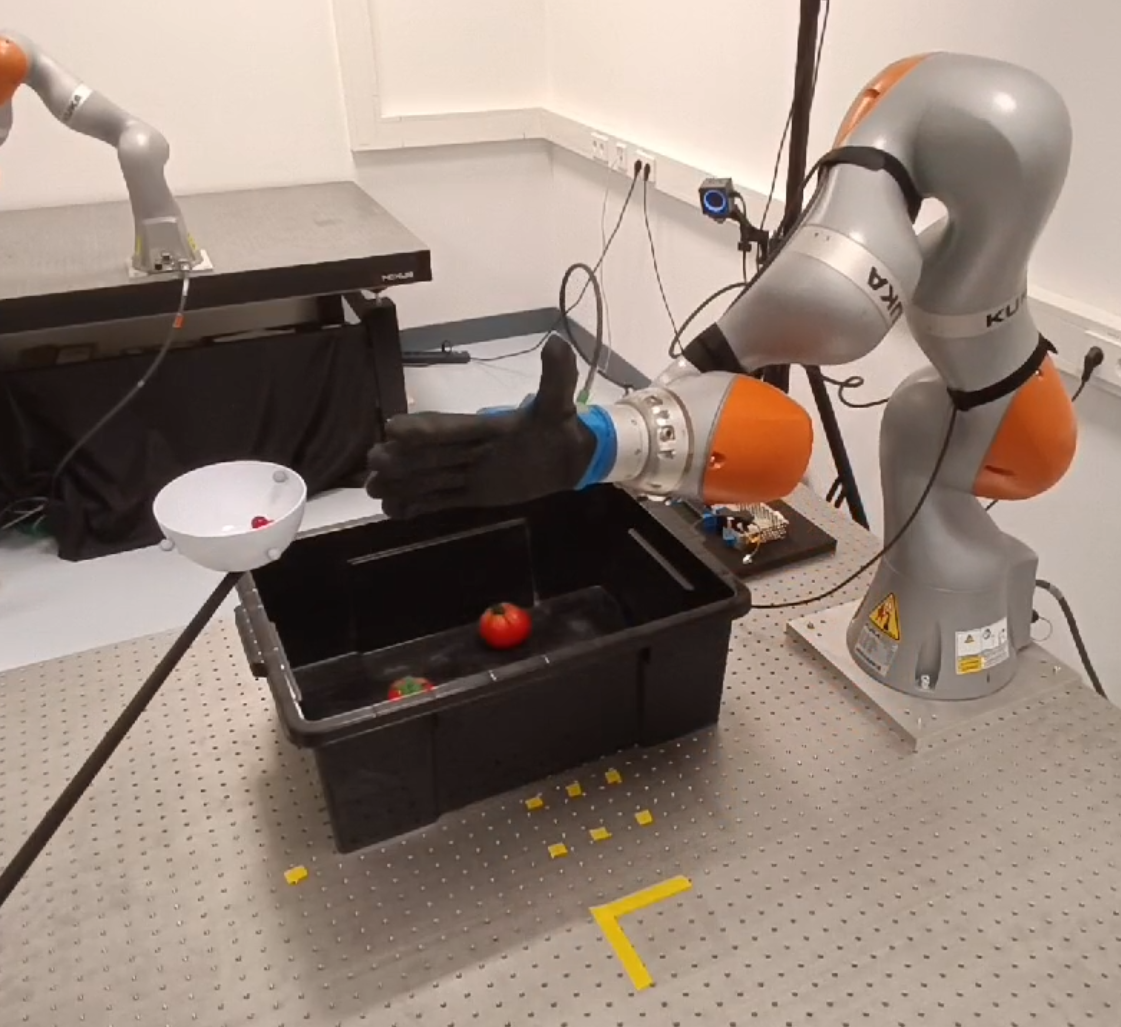}}\label{fig:subfig6}
\hspace{\fill}
\subfigure[\footnotesize{Avoid the bowl}]{\includegraphics[width=0.19\textwidth, keepaspectratio]{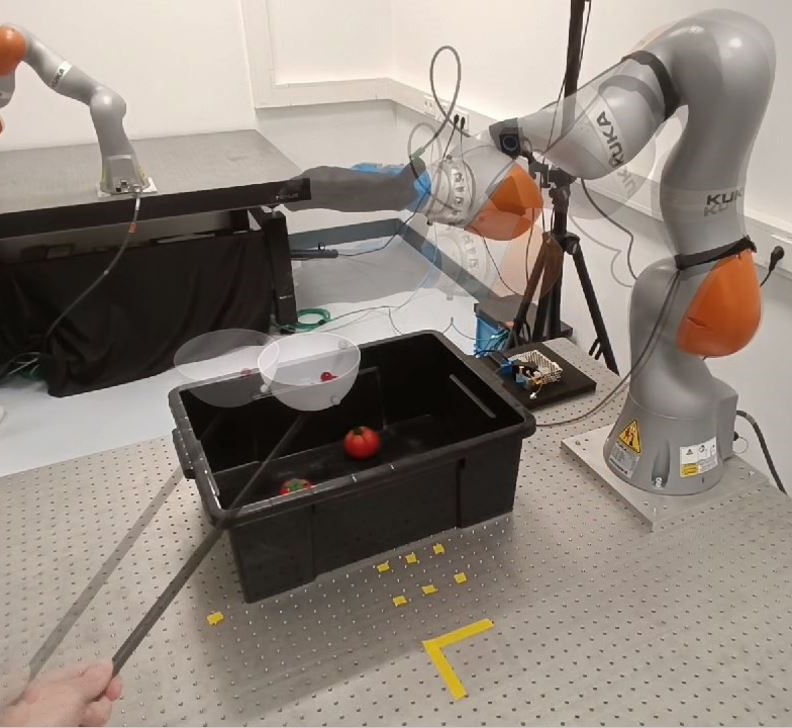}}\label{fig:subfig7}
\hspace{\fill}
\subfigure[\footnotesize{Avoid the hand}]{\includegraphics[width=0.19\textwidth, keepaspectratio]{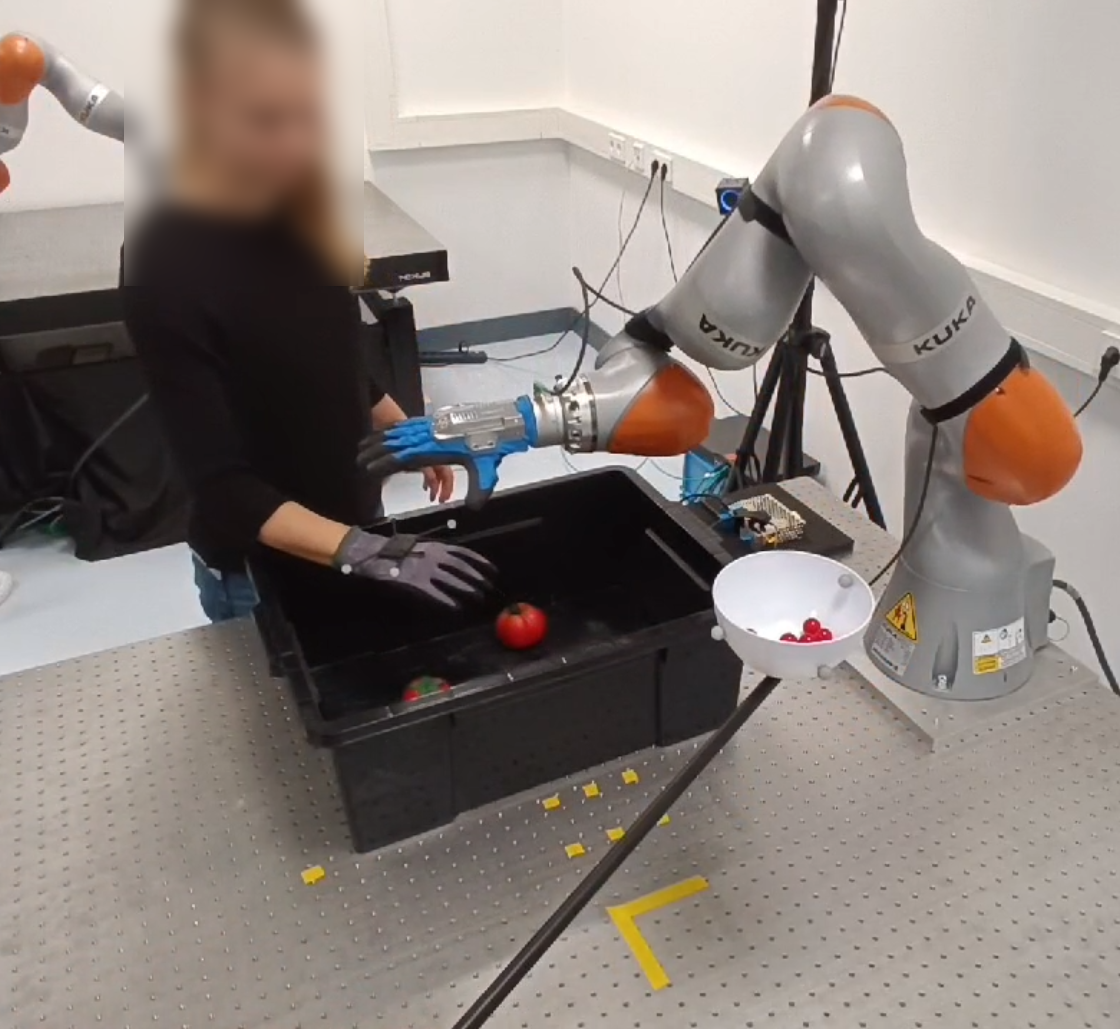}}\label{fig:subfig8}
\hspace{\fill}
\subfigure[\footnotesize{Goal reached}]{\includegraphics[width=0.19\textwidth, keepaspectratio]{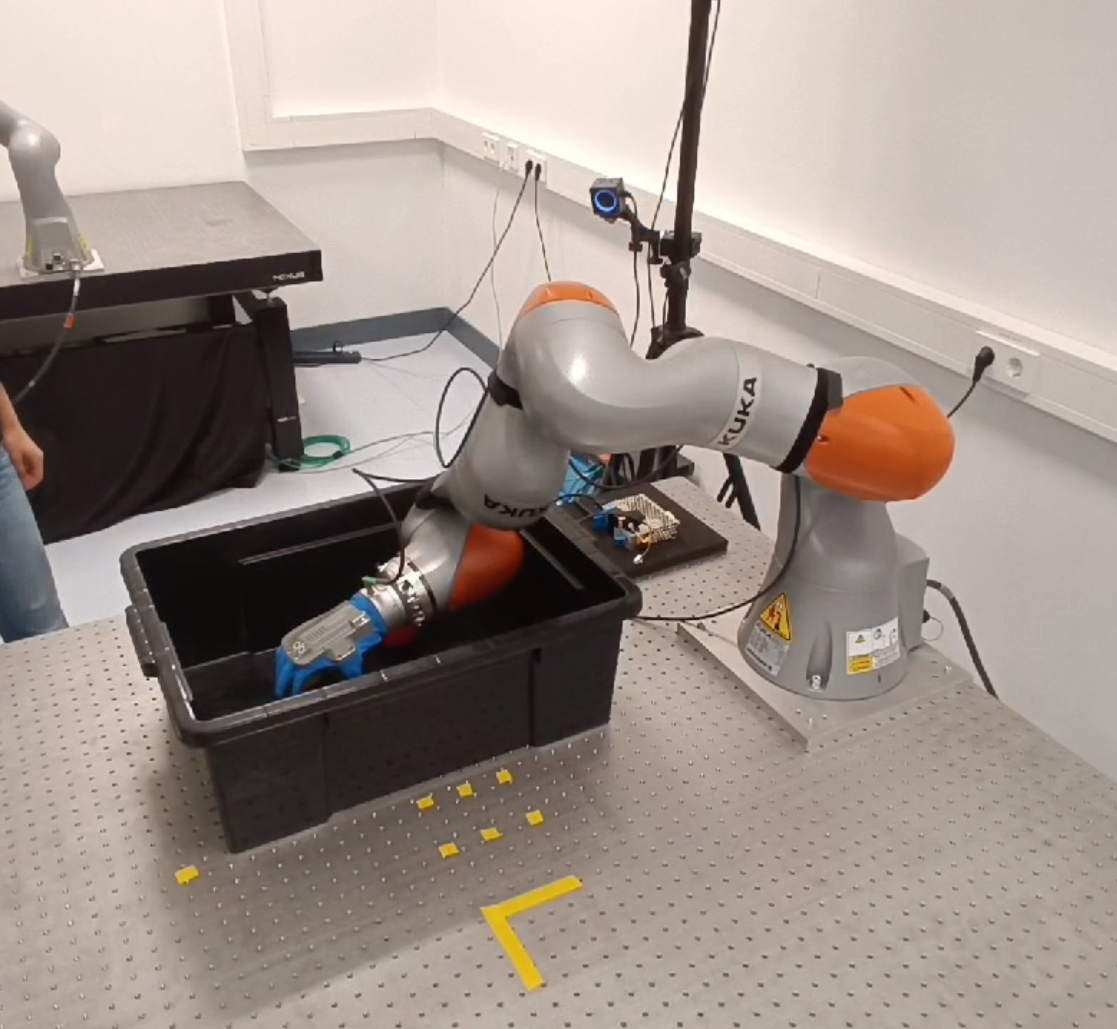}}\label{fig:subfig9}
\caption[]{\footnotesize{Selected time frames of \ac{cpm} during a tomato-picking task with the bowl and hand as dynamic obstacles.}}
\vspace{-2mm}
\label{fig: realworld}
\end{figure*}

\begin{figure*}[t]   
\subfigure[\footnotesize{Initial pose}]{\includegraphics[width=0.19\textwidth, keepaspectratio]{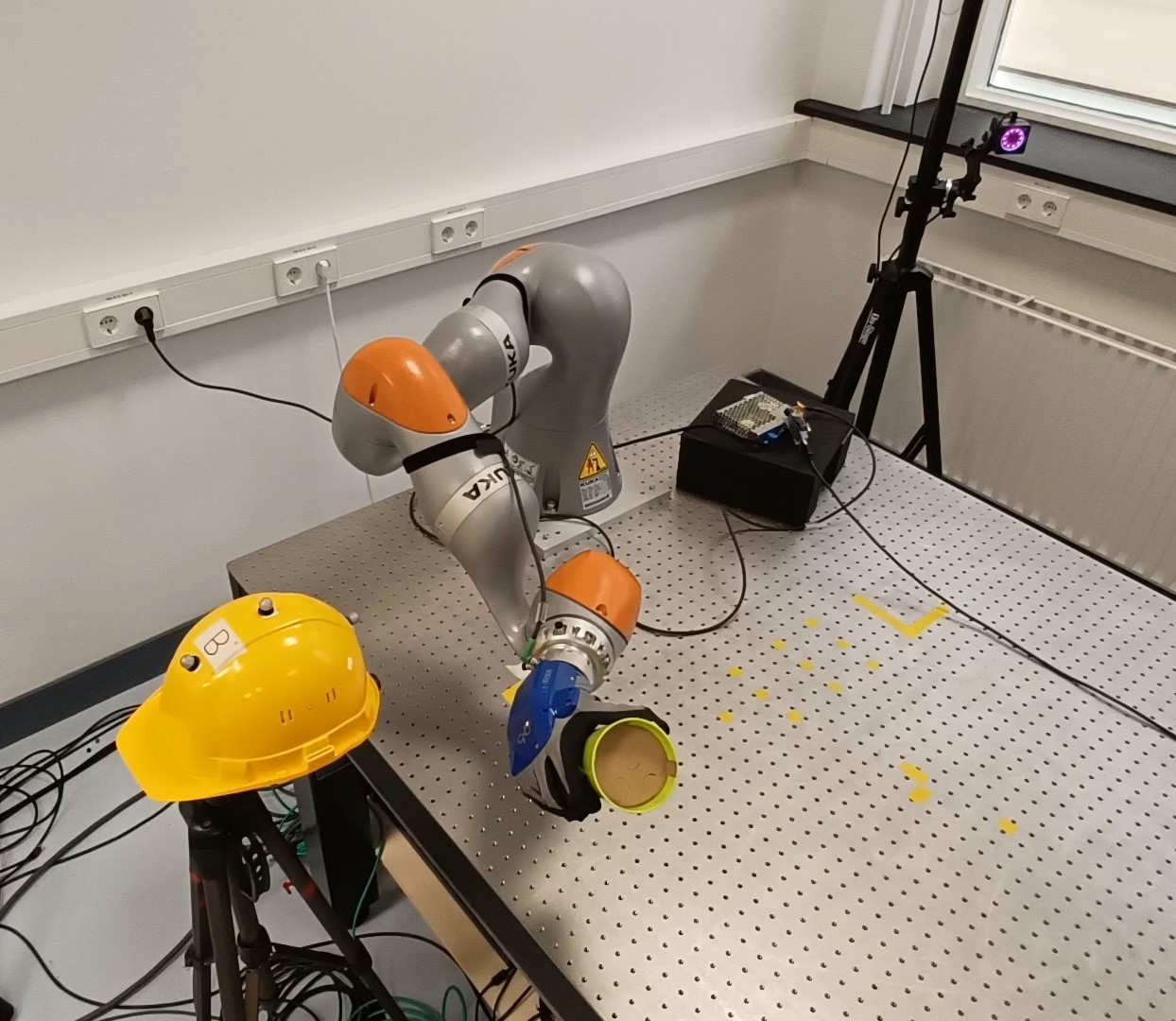}}\label{fig:subfigh5}
\hspace{\fill}
\subfigure[\footnotesize{Avoid the helmet}]{\includegraphics[width=0.19\textwidth, keepaspectratio]{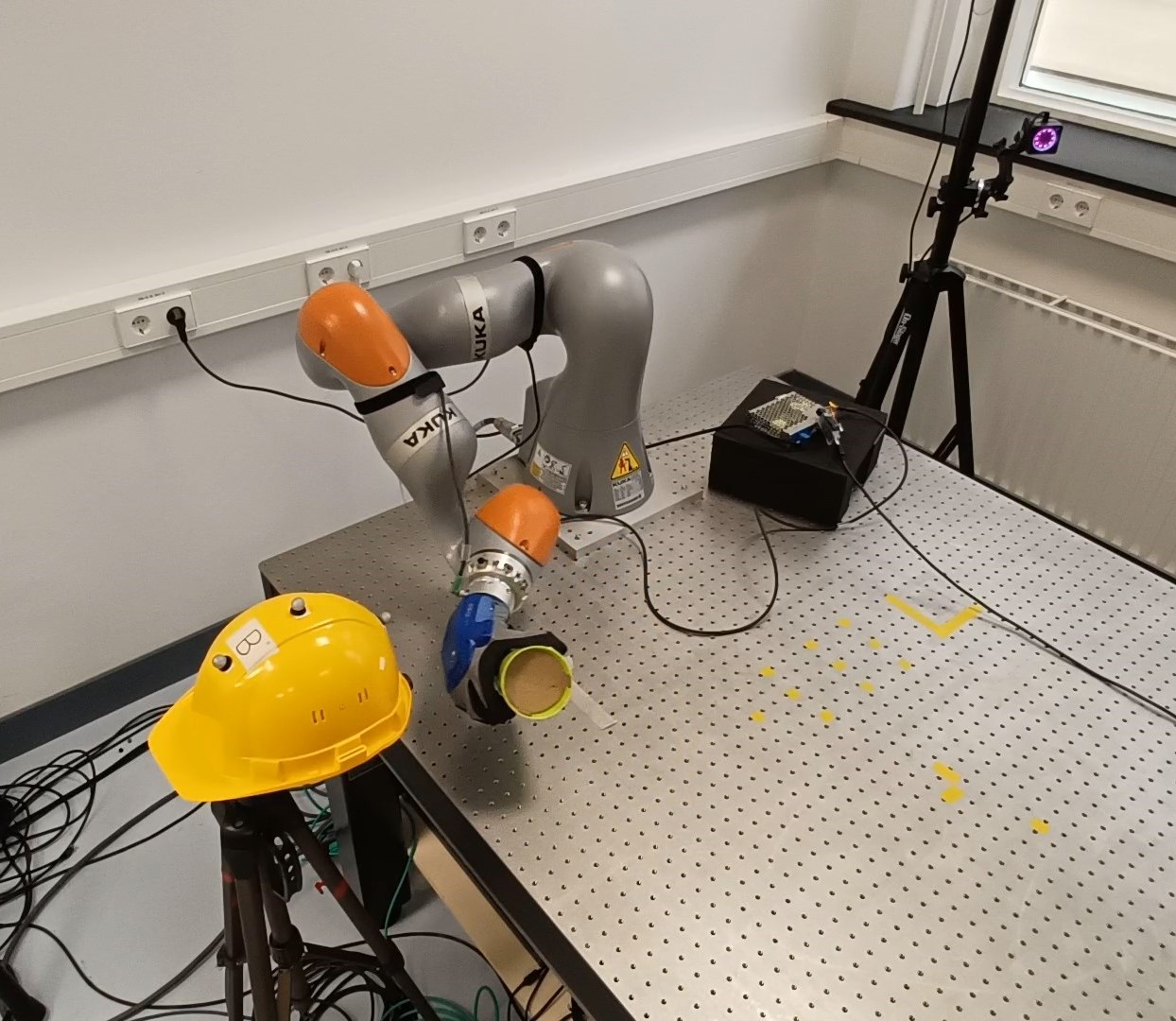}}\label{fig:subfigh6}
\hspace{\fill}
\subfigure[\footnotesize{Avoid the helmet}]{\includegraphics[width=0.19\textwidth, keepaspectratio]{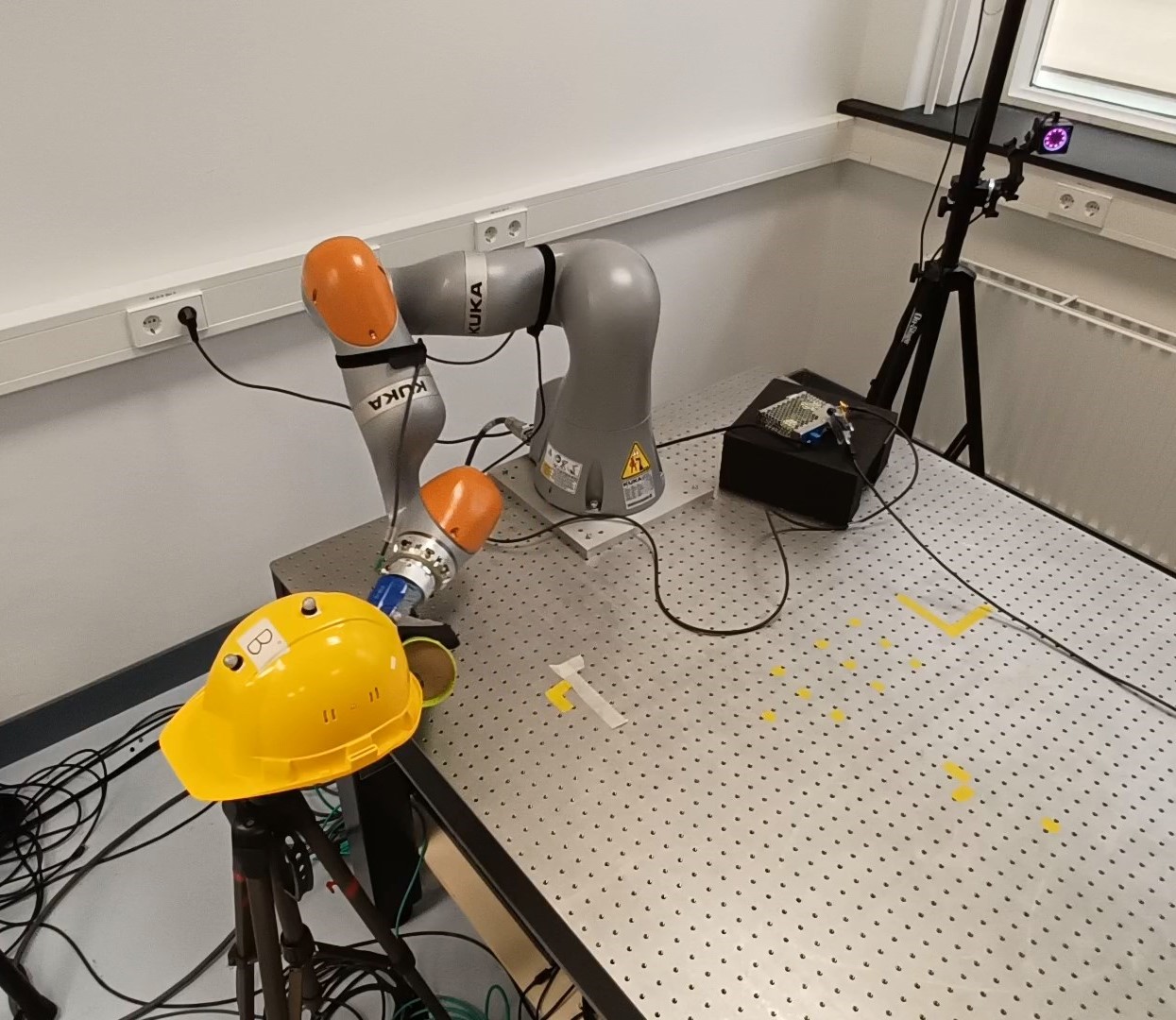}}\label{fig:subfigh7}
\hspace{\fill}
\subfigure[\footnotesize{Avoid the helmet}]{\includegraphics[width=0.19\textwidth, keepaspectratio]{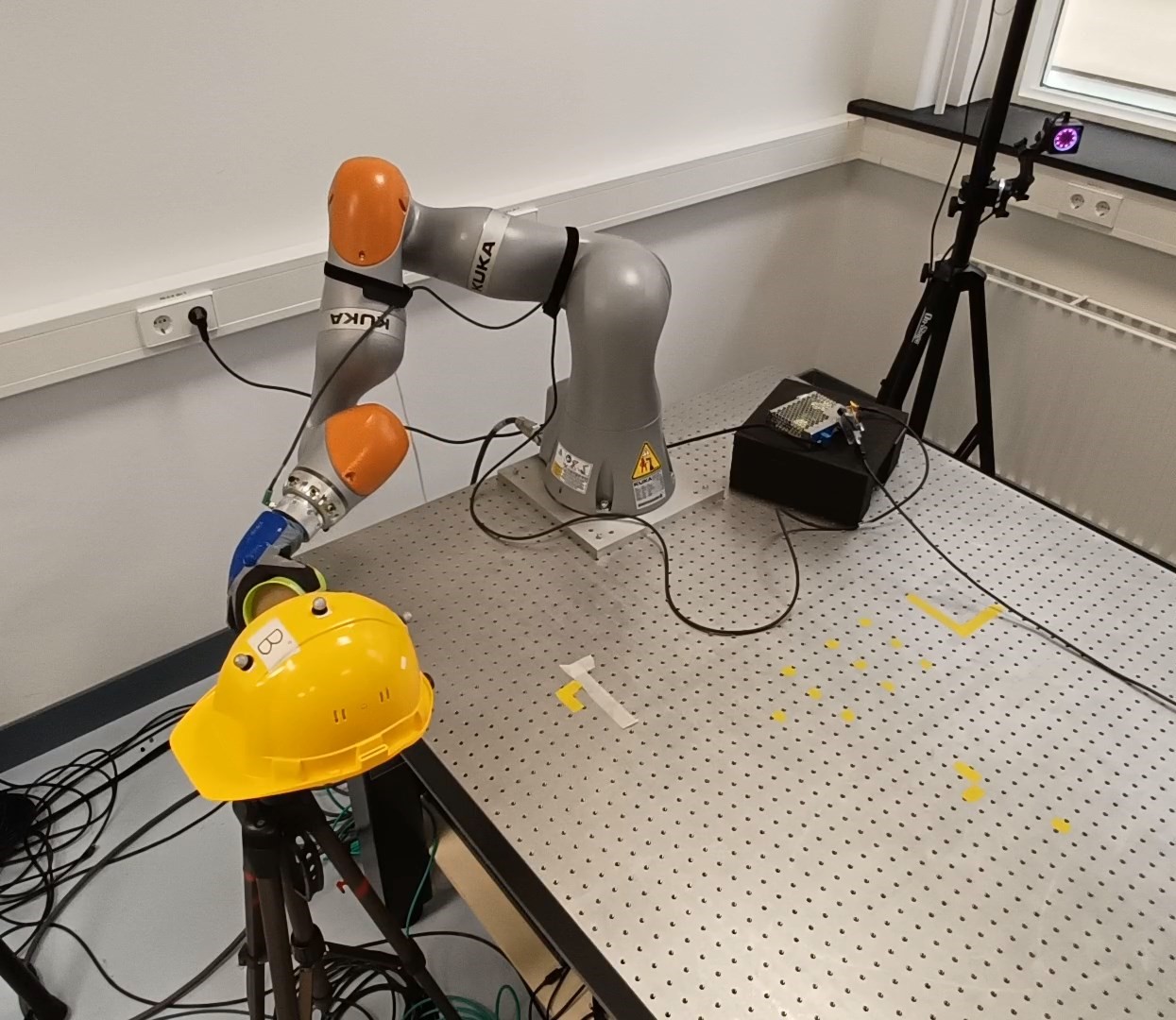}}\label{fig:subfigh8}
\hspace{\fill}
\subfigure[\footnotesize{Goal reached}]{\includegraphics[width=0.19\textwidth, keepaspectratio]{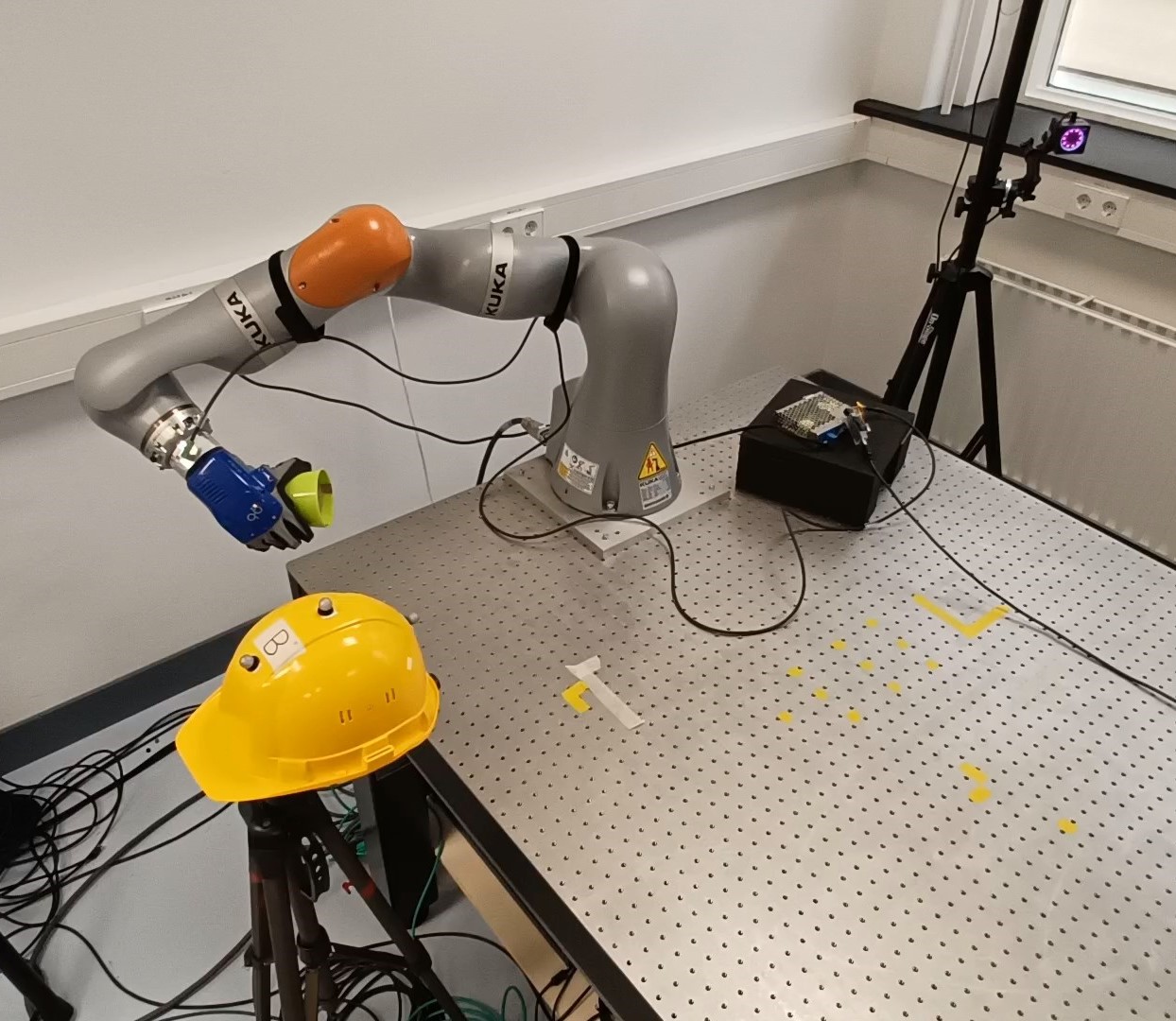}}\label{fig:subfigh9}
\caption[]{\footnotesize{Selected time frames of \ac{cpm} during a pouring task with the yellow helmet as a static obstacle.}}
\vspace{-2mm}
\label{fig: realworld2}
\end{figure*}

\begin{figure*}[th!]   
\subfigure[\footnotesize{Initial pose}]{\includegraphics[width=0.19\textwidth, keepaspectratio]{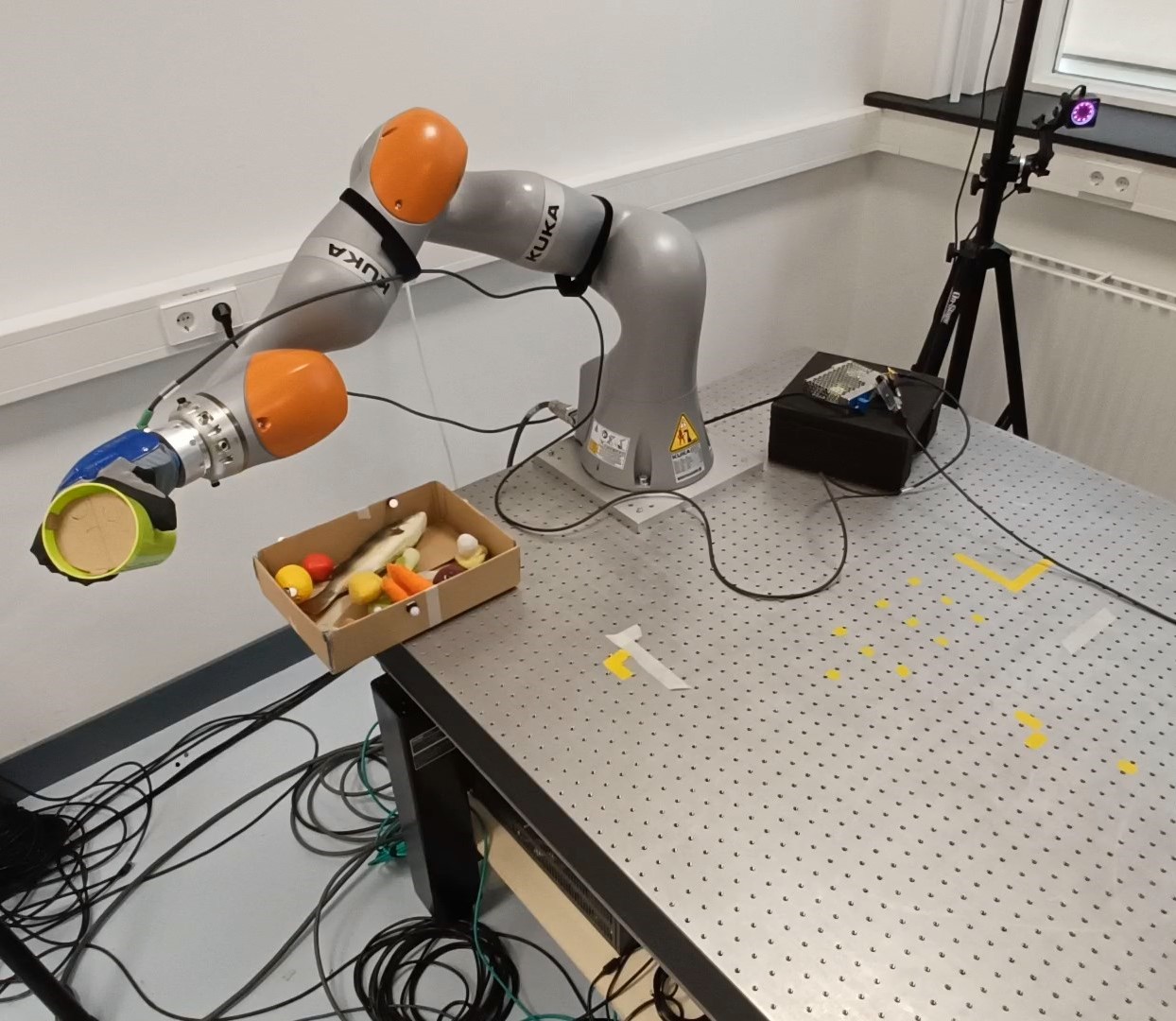}}\label{fig:subfigp5}
\hspace{\fill}
\subfigure[\footnotesize{Move to goal}]{\includegraphics[width=0.19\textwidth, keepaspectratio]{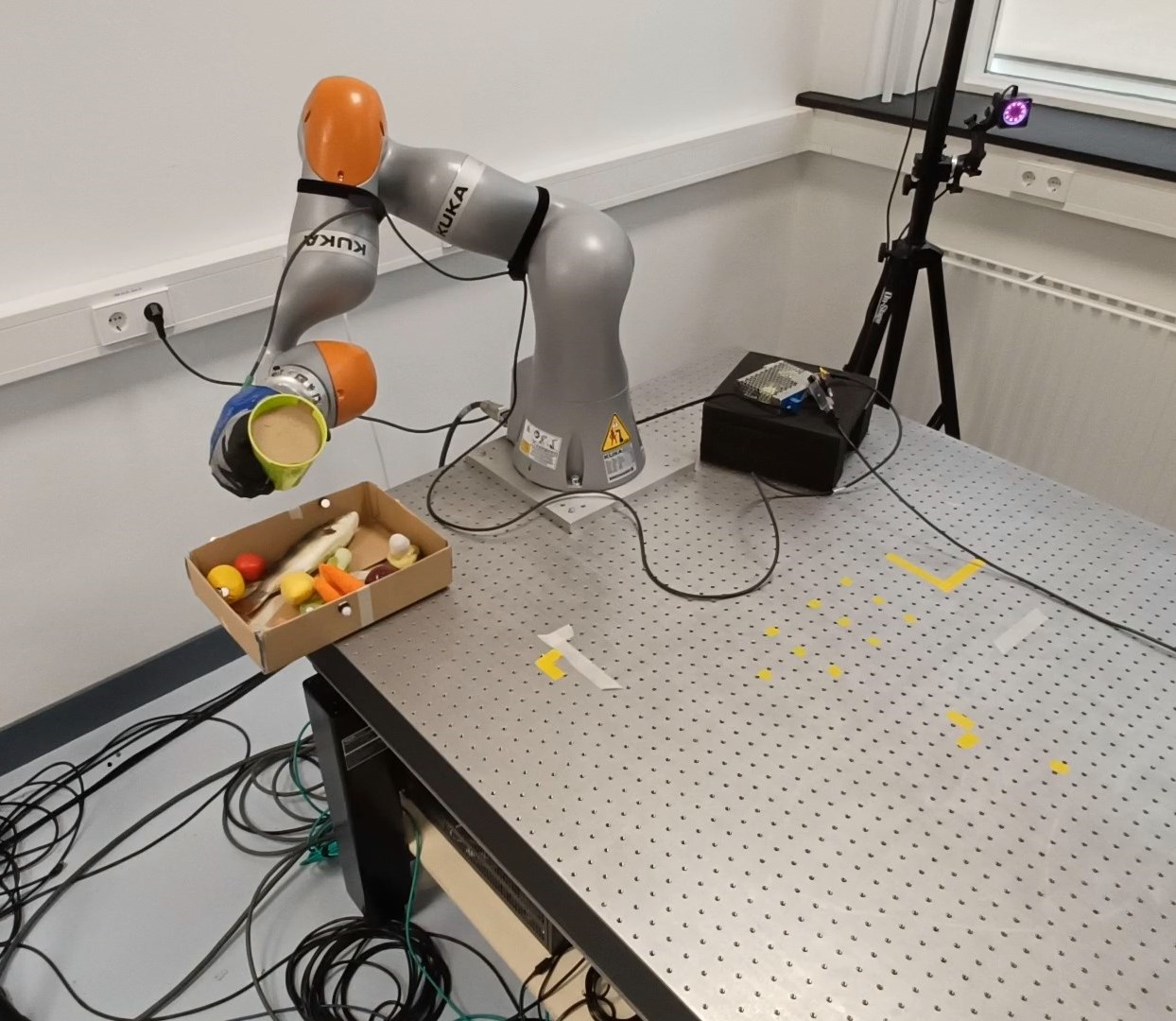}}\label{fig:subfigp6}
\hspace{\fill}
\subfigure[\footnotesize{Goal reached}]{\includegraphics[width=0.19\textwidth, keepaspectratio]{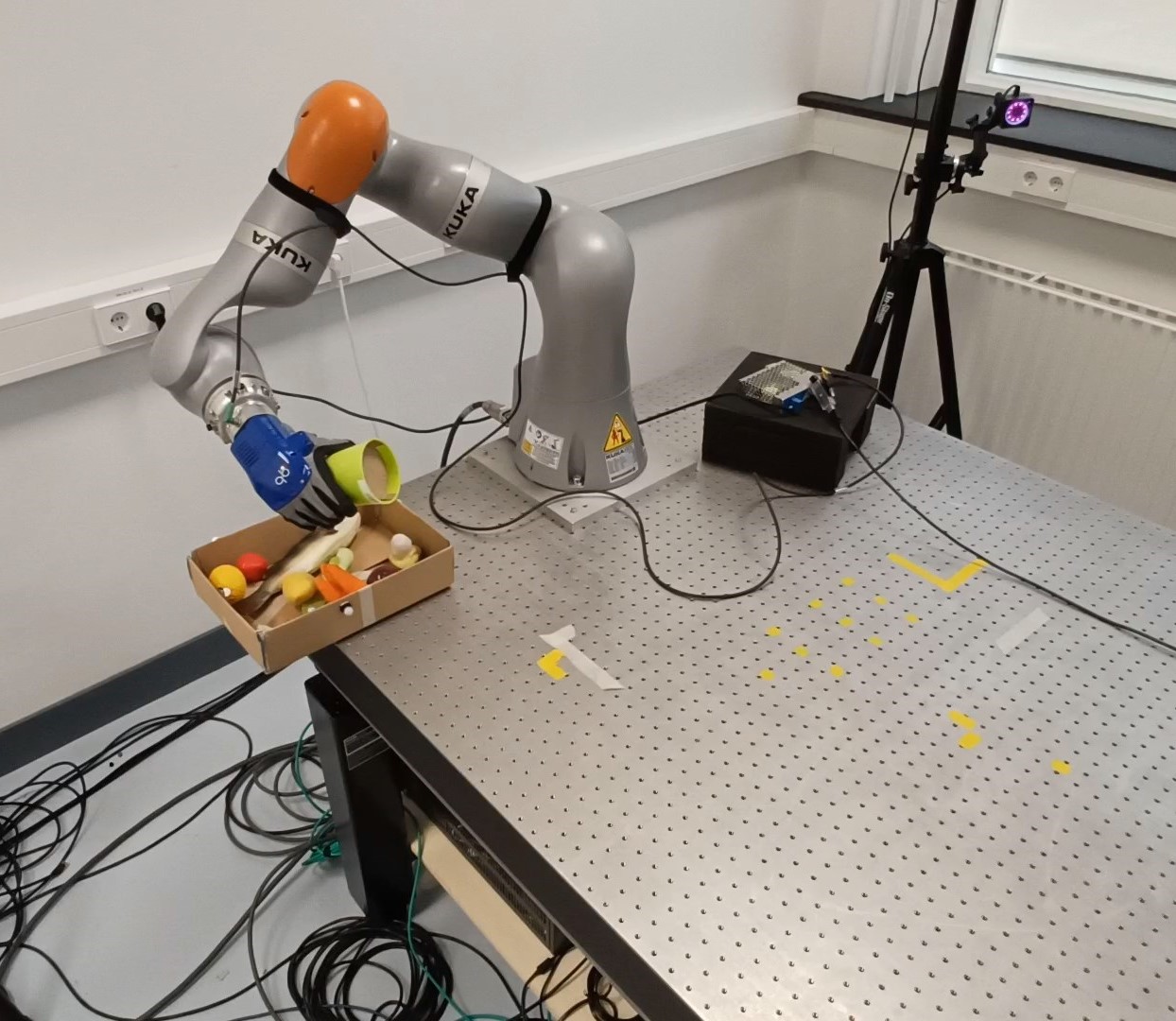}}\label{fig:subfigp7}
\hspace{\fill}
\subfigure[\footnotesize{Goal moved}]{\includegraphics[width=0.19\textwidth, keepaspectratio]{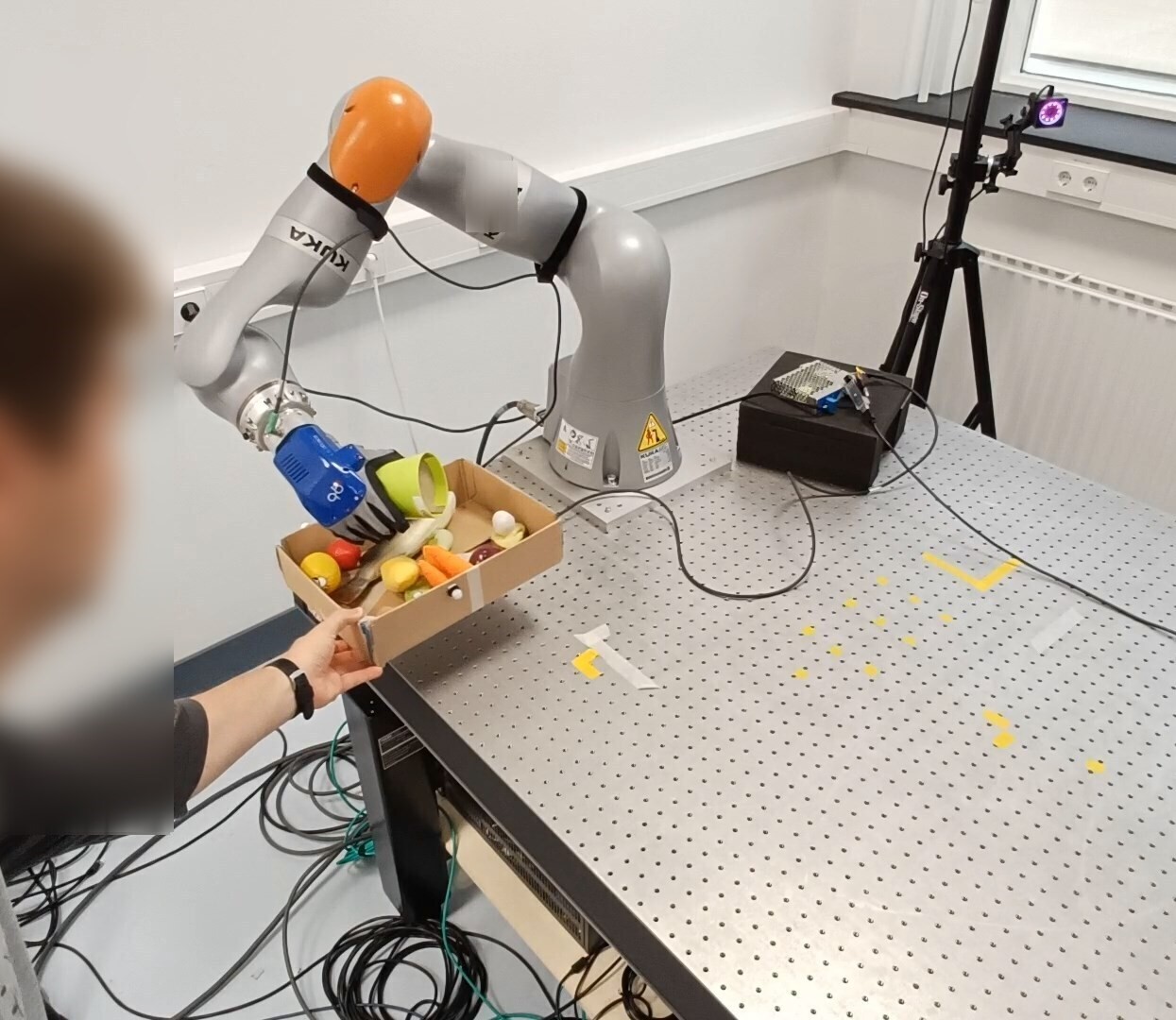}}\label{fig:subfigp8}
\hspace{\fill}
\subfigure[\footnotesize{Goal reached}]{\includegraphics[width=0.19\textwidth, keepaspectratio]{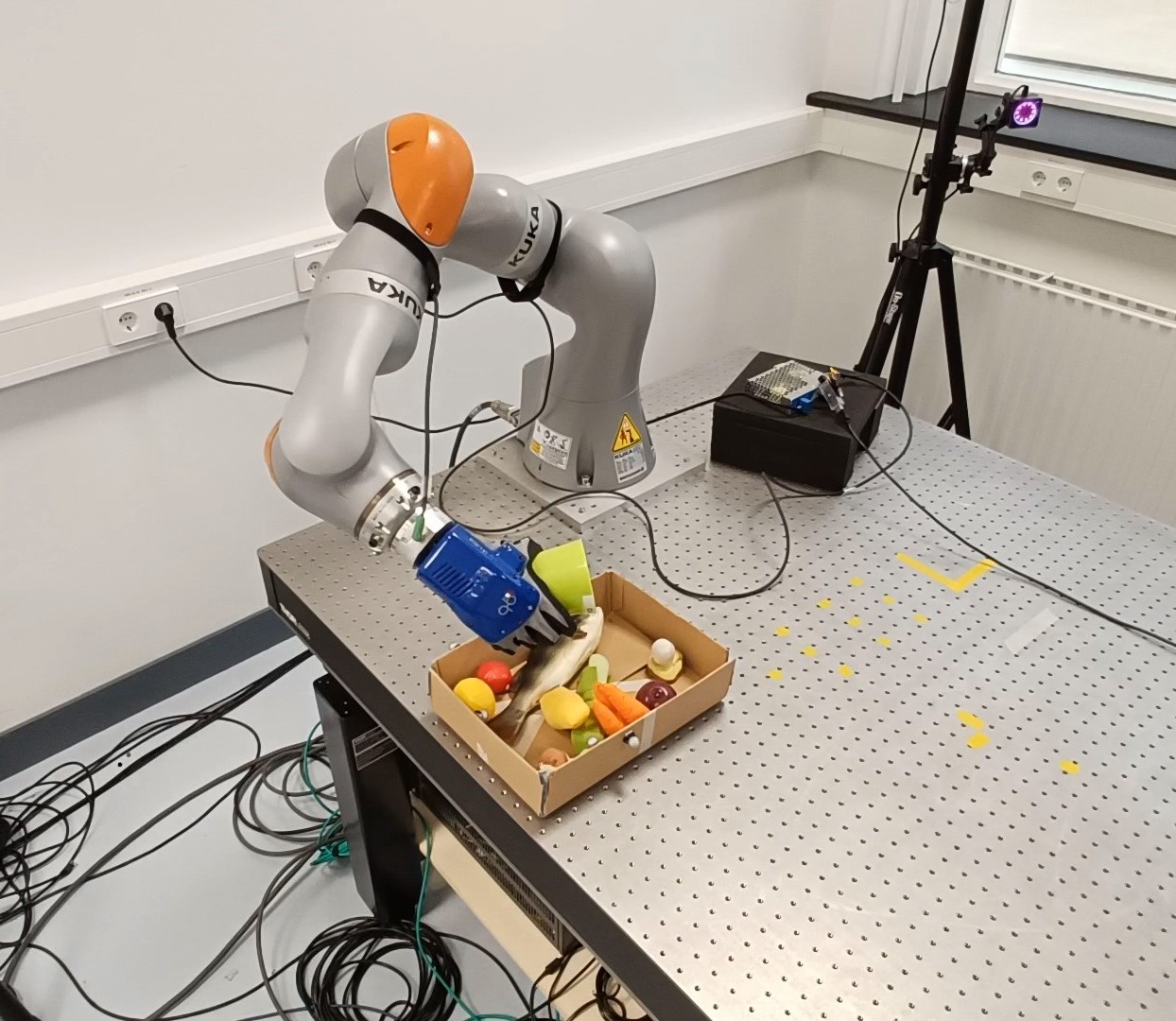}}\label{fig:subfigp9}
\caption[]{\footnotesize{Selected time frames of \ac{cpm} during a pouring task where the goal is changed online by the user.}}
\vspace{-2mm}
\label{fig: realworld3}
\end{figure*}

Experiments are performed on the real 7-DOF for the tomato-picking and pouring task where all obstacles, e.g. a bowl, a person's hand and a helmet, are dynamically tracked in real-time via an optitrack system \SB{and modeled as spheres}. 
In addition to the collision spheres considered during simulation, a collision sphere is added to the collision geometry on the center of the robotic hand with a diameter of 14 $cm$. The desired actions by TamedPUMA are sent at 30 Hz to a joint impedance controller running at 1000 Hz.
Snapshots of a real-world experiment of the \ac{cpm} are illustrated in Fig.~\ref{fig: realworld}, Fig.~\ref{fig: realworld2} and Fig~\ref{fig: realworld3}, and the videos in the attached material show several movements for the simulated and real-world experiments \SB{using \ac{fpm} and \ac{cpm}}. 
If the obstacles are not blocking the trajectory of the robot, the observed behavior of the proposed methods, \ac{fpm} and \ac{cpm}, are similar to \ac{puma} and showcases clearly the learned behavior as demonstrated by the human. The user can push the robot away from the goal or change the goal online (Fig.~\ref{fig: realworld3}), and TamedPUMA recovers from this disturbance.
In the presence of obstacles, \ac{fpm} and \ac{cpm} achieve collision avoidance between the considered links on the robot and the obstacles while reaching the goal pose, as illustrated in Fig.~\ref{fig: realworld}, and Fig.~\ref{fig: realworld2}.

A low-level joint impedance controller tracks the desired velocities and positions, outputting torque commands. This might decrease tracking performance compared to the simulated experiments, but allows users to push the robot away. If the user is no longer applying force to the robot, the system will converge to the goal pose, as illustrated in the videos in the attached material.

%% file: sections/6_conclusion.tex
\section{Conclusion}
\label{sec:conclusion}
Imitation learning via stable motion primitives is a suitable approach for learning motion profiles from demonstrations while providing convergence to the goal. We introduced TamedPUMA, a safe and stable extension of learned stable motion primitives augmented with the recently developed geometric fabrics for safe and stable operations in the presence of obstacles.
We proposed two variations, the Forcing Policy Method and Compatible Potential Method, ensuring respectively that the goal is stable, or the stronger notion that the system converges towards the reachable goal.
Experiments were carried out both in simulation and in the real world.
When trained on a tomato-picking task or pouring task, the proposed TamedPUMA generates a desired motion profile using a \ac{dnn} while taking whole-body collision avoidance and joint limits into account, with a computation time of just 4-7~$ms$.